\documentclass[prb,twocolumn,superscriptaddress,showpacs]{revtex4-1}

\usepackage{hyperref}
\usepackage{rotating}
\usepackage{graphicx}
\usepackage{latexsym}
\usepackage{amssymb}
\usepackage{amsfonts}
\usepackage{soul}
\usepackage{color}
\usepackage{amsmath}
\usepackage{color}
\usepackage{lipsum}
\usepackage{natbib}

\begin{document}

\title{Multiband mechanism of the pair fluctuation screening}

\author{T. T. Saraiva}
\affiliation{National Research University Higher School of Economics, 101000, Moscow, Russia}

\author{A. A. Shanenko}
\affiliation{National Research University Higher School of Economics, 101000, Moscow, Russia}

\author{A. Vagov}
\affiliation{Institut f\"{u}r Theoretische Physik III, Bayreuth Universit\"{a}t, Bayreuth 95440, Germany}

\author{A. S. Vasenko}
\affiliation{National Research University Higher School of Economics, 101000, Moscow, Russia}
\affiliation{I.E. Tamm Department of Theoretical Physics, P.N. Lebedev Physical Institute, Russian Academy of Sciences, 119991 Moscow, Russia}

\date{\today}
\begin{abstract}
Recent chain-like structured materials have shown a robust superconducting phase. These materials exhibit the presence of quasi-one-dimensional bands (q1D) coupled to conventional higher-dimensional bands. On the mean-field level such systems have a high critical temperature when the chemical potential is close to the edge of a q1D band and the related Lifshitz transition is approached. However, the impact of the pair fluctuations compromises the mean-field results. Recently it has been demonstrated that these fluctuations can be suppressed (screened) by a specific multiband mechanism based on the pair-exchange coupling of the q1D condensate to a stable higher-dimensional one. In the present work we demonstrate that strikingly enough, this mechanism is not very sensitive to the basic parameters of the stable condensate such as its strength and dimensionality. For example, even the presence of a passive higher-dimensional band, which does not exhibit any superconducting correlations when taken as a separate superconductor, results in suppression of the pair fluctuations.    
\end{abstract}

\maketitle
\section{Introduction}

Since their first experimental detection in MgB$_2$,~\cite{Nagamatsu2001,Larbalestier2001} multiband superconductors have shown a rich phenomenology improving our understanding and knowledge of superconductivity~\cite{Orlova2013,Milosevic2015,Huang20}.
The fundamental difference of the multiband superconductors from the conventional single-band superconducting materials is that the interference of multiple contributing condensates can result in significant deviations from the single-condensate physics.
Recently it has been demonstrated that such interference affects the superconducting fluctuations, leading to the multiband fluctuation screening mechanism.~\cite{Salasnich2018,Saraiva2020} The pair exchange coupling between the multiple condensates can wash out the fluctuations of the order parameter and thus amplify the critical temperature of the system.
It has been revealed that the severe fluctuations of the quasi-one-dimensional (q1D) condensate are suppressed by an almost negligible pair-exchange coupling to the stable BCS condensate.~\cite{Saraiva2020} However, it was not investigated how the multiband screening mechanism depends on the parameters of the stable higher-dimensional condensate such as its strength and dimensionality. Here, motivated by on-going experiments with the multiband q1D superconductors A$_2$Cr$_3$As$_3$ (A = K, Rb, Cs)~\cite{Bao2015,Tang2015A,Wu2019} and similar organic materials \cite{Wang2017a,Wang2017b,Wang2017c}, we are going to fill this gap. 
As a prototype of chain like structured multiband superconducting materials, we consider a two-band superconductor with a q1D band coupled to a stable 2D/3D condensate and investigate the dependence of the fluctuation shifted critical temperature on the system parameters. 

\section{Theoretical Approach}\label{secII}

\begin{figure}[!ht]
\centering
\includegraphics[width=\linewidth]{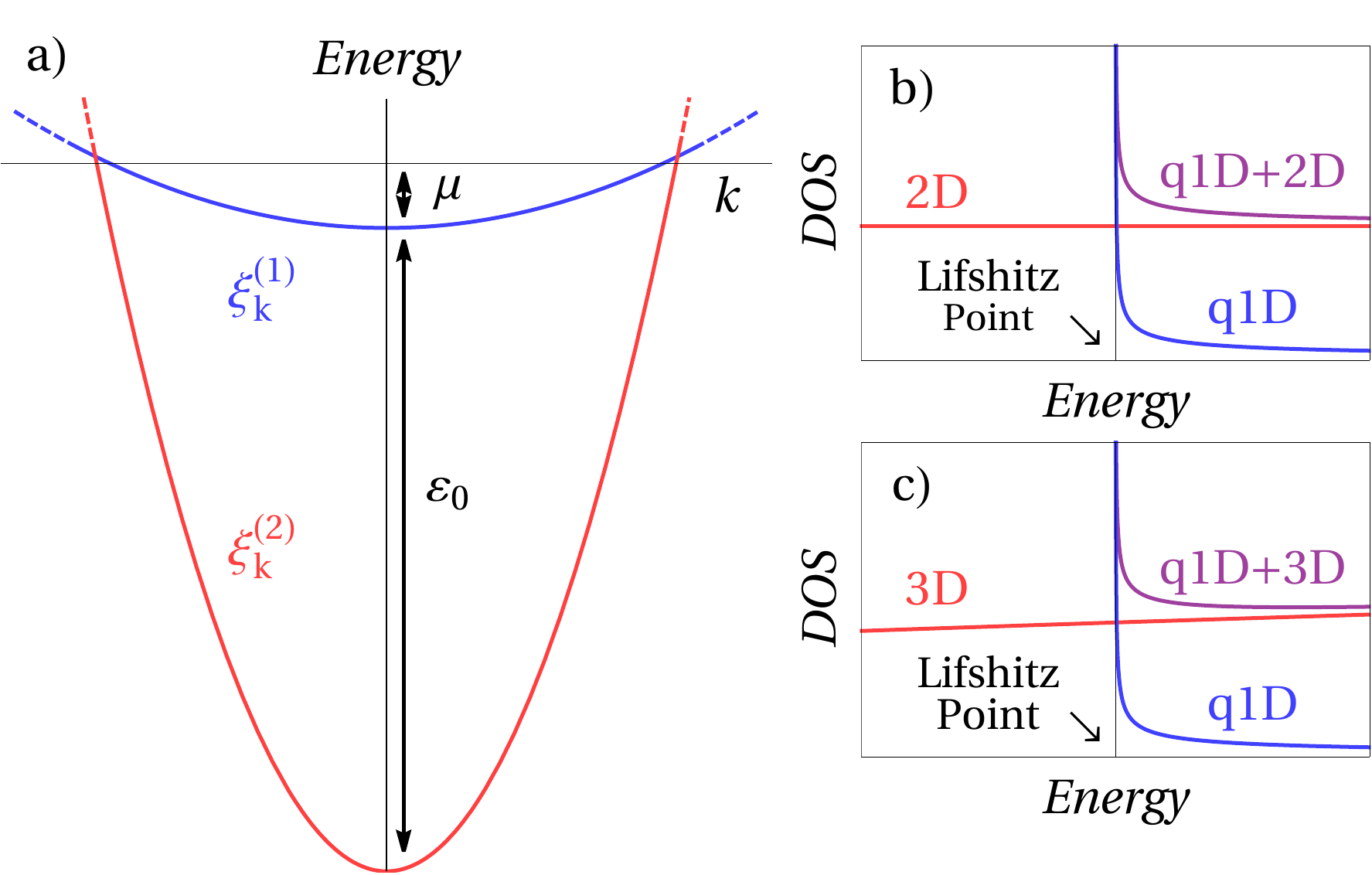}
\caption{\label{sketch}a) Sketch of the dispersion relations of the shallow and deep bands, $\xi_k^{(1)}$ and $\xi_k^{(2)}$ respectively. The distance between the bottom of the bands is $\varepsilon_0$. To the right, in plots b) and c) we show sketches of the DOSs in the 2D and in the 3D cases in the deep band regime, respectively, combined to the q1D band. Blue and Red (purple) lines represent partial (total) DOS.}
\end{figure}

We consider the standard multiband generalization of the BCS model~\cite{Suhl1959,Moskalenko1959} with a pair exchange coupling between the two contributing bands. The coupling matrix $g_{\nu\nu'}$ ($\nu,\nu'=1,2$) is symmetric, where $\nu=1$ stays for the q1D band and $g_{11}>g_{22},g_{12}$, e.g., the q1D band is a stronger one. The weaker band corresponds to $\nu=2$ and, taken in its passive limit, it has $g_{22}=0$. In this case, the second gap function is nonzero due to the pair exchange coupling between bands $1$ and $2$. We choose the spherical Fermi surfaces for both bands with the dispersions (absorbing the chemical potential $\mu$) 
\begin{equation}
\qquad\xi_{\bf k}^{(1)}=\frac{\hbar^2 k_z^2}{2m_1}-\mu\quad\mbox{and}\quad\xi_{\bf k}^{(2)}=\varepsilon_0+\frac{\hbar^2\bf k^2}{2m_2}-\mu,
\label{eq.xinu}
\end{equation}
where $m_{1,2}$ are the effective electronic masses for each band, the q1D single-electron energy varies only in the z-direction (depending on $k_z$) and the single-electron energy in the higher-dimensional band depends on the 2D/3D wavevector ${\bf k}$. The condensate in band $2$ is stable so that the band is deep enough with $\varepsilon_0<0$ and $|\varepsilon_0|\gg\mu$, as sketched in Fig.~\ref{sketch}.

The Hamiltonian reads
\begin{align}
\mathcal{H}=&\int d^3{\bf r}\Bigg\{\sum_{\nu=1,2} \Bigg[\sum_{\sigma=\uparrow,\downarrow}\hat{\psi}_{\nu\sigma}^\dagger({\bf r})T_\nu({\bf r})\hat{\psi}_{\nu\sigma}({\bf r})\nonumber\\
&+\left(\hat{\psi}_{\nu\uparrow}^\dagger({\bf r})\hat{\psi}_{\nu\downarrow}^\dagger({\bf r}) \Delta_\nu({\bf r}) + {\rm h.c.}\right)\Bigg] + \langle \vec\Delta, \check{g}^{-1}\vec\Delta \rangle \Big\}
\label{eq.H}
\end{align}
where ${\hat\psi}^{\dagger}_{\nu\sigma}({\bf r})$ and ${\hat\psi}_{\nu\sigma}({\bf r})$ are the field operators for the carriers in band $\nu$, $T_{\nu}({\bf r})$ is the single-particle Hamiltonian corresponding to with the single-particle energies given by Eq.~(\ref{eq.xinu}), and $\Delta_{\nu}({\bf r})$ is the gap function for band $\nu$. We also use a vector notation $\vec\Delta = (\Delta_1,\Delta_2)$ with $\langle.,.\rangle$ the corresponding inner product, and $\check{g}^{-1}$ is the inverse of the coupling matrix. 

The Hamiltonian is solved together with the self-consistency equation written in terms of the anomalous Green functions $R_\nu({\bf r}) = \left<\psi_{\nu\uparrow}(\bf r)\psi_{\nu \downarrow}(\bf r)\right>$ as
\begin{equation}
\Delta_\nu({\bf r})=\sum_{\nu'=1,2}g_{\nu\nu'}R_{\nu'}({\bf r}).\label{eq.selfcons}
\end{equation}
From Eqs.~(\ref{eq.H}) and (\ref{eq.selfcons}) one derives \cite{Shanenko2011,Vagov2012EGL} the linearized gap equation
\begin{equation}
\sum_{\nu'=1,2}\gamma_{\nu\nu'}\Delta_{\nu'}=\mathcal{A}_\nu\Delta_\nu\quad\Rightarrow\quad\breve{L}\vec{\Delta}=0,\label{eq.gapex}
\end{equation}
where the auxiliary matrix $L_{\nu\nu'}=\gamma_{\nu\nu'}-\mathcal{A}_\nu\delta_{\nu\nu'}$ is introduced in terms of the inverse coupling matrix $\breve{\gamma}=\breve{g}^{-1}$ and the coefficients $\mathcal{A}_{\nu}$ given by (see Appendix~\ref{SecApp}):
\begin{align}
&\mathcal{A}_{1}=N_1\int\limits_{-\tilde\mu}^{1}dy\frac{\tanh(y/2\tilde{T}_{c0})}{y\sqrt{y+\tilde\mu}}\label{eq.dos1}\\
&\mathcal{A}_2=N_2\ln\left(\frac{2e^\Gamma}{\pi\tilde{T}_{c0}}\right),
\end{align}
where $\Gamma\approx0.577$ is the Euler-Mascheroni constant, $T_{c0}$ is the mean-field critical temperature of the system and quantities marked by a tilde are normalized by the cutoff energy $\hbar\omega_c$. The parameter $N_1=\sigma^{(xy)}\sqrt{m_z/8\pi^2\hbar^2}$ has units of DOS, but the divergent part is kept inside the integral as shown in Eq.~(\ref{eq.dos1}). The term $\sigma^{(xy)}$ accounts for the DOS in the $x$ and $y$ directions. For the higher-dimensional band we have $N_2^{(2D)}=\sigma^{(x)}m_2/2\hbar^2$ ($\sigma^{(x)}$ accounts for the DOS in the $x$ direction) in the 2D case and $N_2^{(3D)}=m_2k_F/2\pi^2\hbar^2$ in the 3D case. The DOSs are sketched in Fig.~\ref{sketch}.

\begin{figure}[t]
\centering
\includegraphics[width=0.8\linewidth]{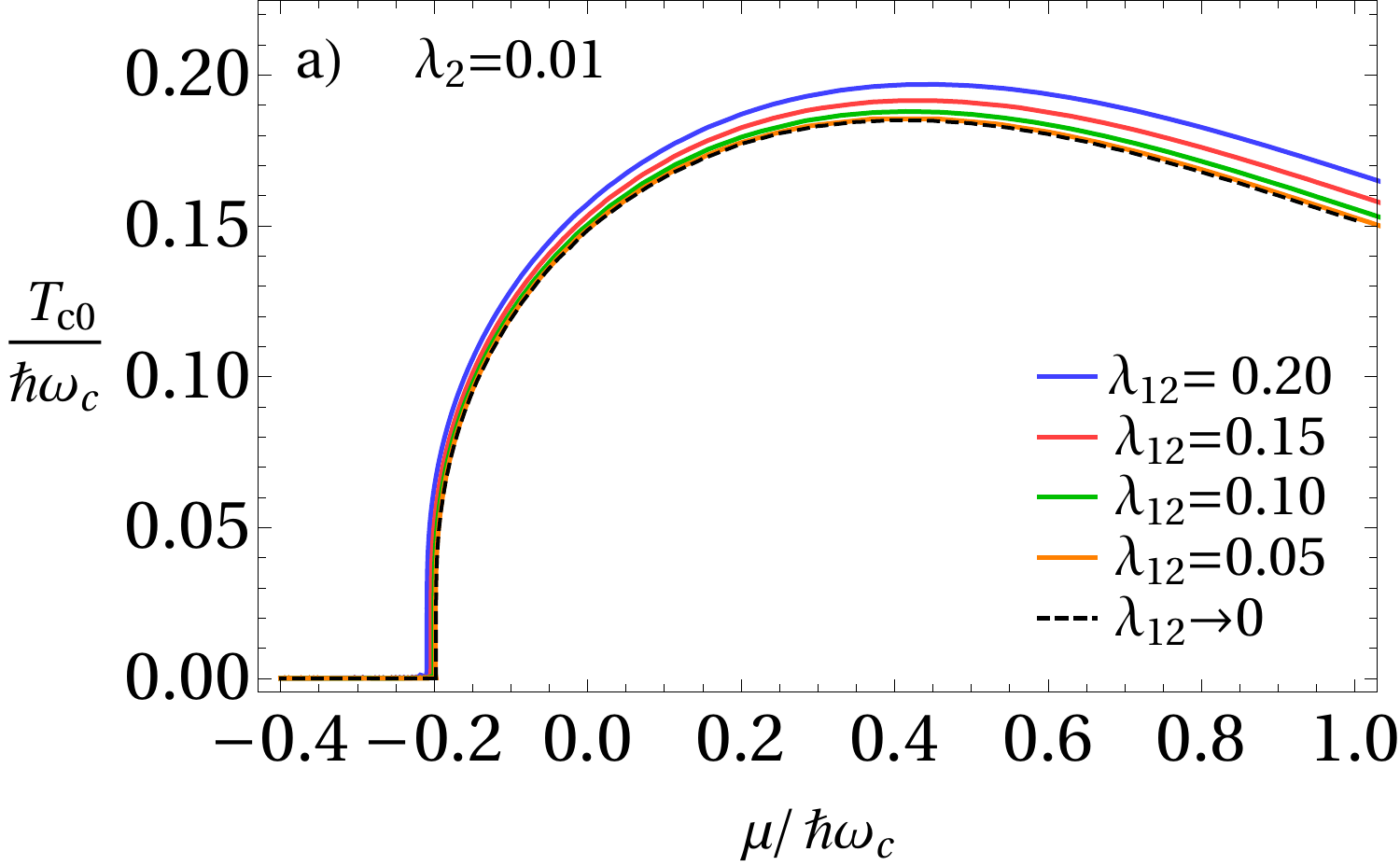}\\
\includegraphics[width=0.8\linewidth]{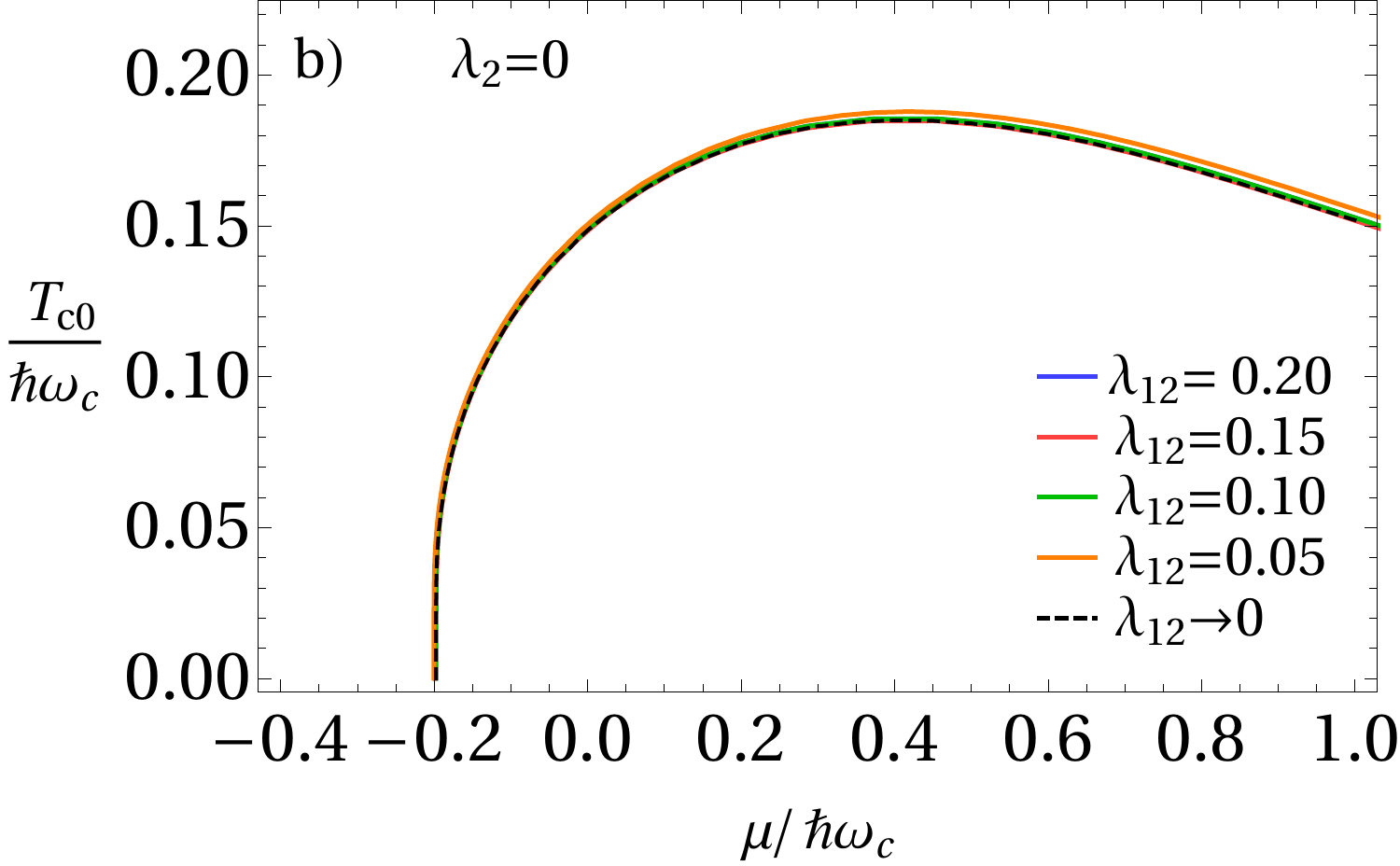}
\caption{\label{fig.tc0}
Mean-field critical temperature, $T_{c0}$, as function of the chemical potential (both quantities are expressed in units of the Debye energy). In plot a) the weaker band has coupling $\lambda_2=0.01$, which is  The dimensionless coupling of the stronger band is $\lambda_1=0.2$ and we used different values of interband coupling $\lambda_{12}$, shown in the figures. The dashed line represents $T_{c0}$ in the limit of uncoupled bands, $\lambda_{12}\to0$.}
\end{figure}

In fact, the couplings and the partial DOSs can be combined in a smaller set of parameters when one expresses the system in terms of the dimensionless couplings
\begin{equation}
\lambda_1=g_{11}N_1,\quad
\lambda_2=g_{22}N_2,\quad
\lambda_{12}=g_{12}\sqrt{N_1N_2}.
\end{equation}
From Eq.~(\ref{eq.gapex}), one can obtain the equation for the mean-field critical temperature of the two-band system, $T_{c0}$:
\begin{align}\label{eq.Tc02b}
&\left(g_{11}-G\mathcal{A}_1\right)\left(g_{22}-G\mathcal{A}_2\right)-g_{12}^2=0,\\
&\mbox{or}\nonumber\\
&\left(\lambda_1\mathcal{A}_1-1\right)\left(\lambda_2\mathcal{A}_2-1\right)-\lambda_{12}^2=0,
\end{align}
where $G=g_{11}g_{22}-g_{12}^2$. Taking the highest from both solutions of Eq.~(\ref{eq.Tc02b}) as the critical temperature of the system, it becomes bounded from bellow by the solution of the isolated single-band systems.
The solutions for $T_{c0}$ as function of the chemical potential (both in units of $\hbar\omega_c$) for different interband couplings are displayed in color plots of Fig.~\ref{fig.tc0}.
By choosing the q1D band as the strongest band, the effect of introducing the second (deep) band is to increase $T_{c0}$ but very weakly for $\lambda_1\gg\lambda_2\to0$.

The contribution of fluctuations to the critical temperature can be obtained from the GL free energy. This is done by considering the order parameter as a linear combination of the eigenvectors of the matrix $\breve{L}$
\begin{equation}
\vec{\eta}_+=
\left(\begin{array}{c}
1\\
S
\end{array}\right),\qquad\vec{\eta}_-=
\left(\begin{array}{c}
-S\\
1
\end{array}\right),
\end{equation}
where $S=g_{12}\mathcal{A}_1=\lambda_{12} \mathcal{A}_1/\chi^{1/2}$. The equation $\breve{L}\vec{\Delta}=0$ states that $\breve{L}$ must have at least one null eigenvector. Considering the non-degenerate case, the eigenvector $\vec{\eta}_+$ must be such that
\begin{equation}
\vec{\Delta}=\Psi({\bf r})\vec{\eta}_+.
\end{equation}
The function $\Psi({\bf r})$ is the GL order parameter of the system and it obeys the single-component GL equation (see Appendix~\ref{SecApp} and Refs.~\onlinecite{Salasnich2018,Saraiva2020})
\begin{equation}
a\Psi+b\Psi^3+\sum\limits_{i=x,y,z}\mathcal{K}_i\nabla_i^2\Psi=0
\end{equation}
with the coefficients given by
\begin{align}
&a=a_{1}+a_{2}S^2,\\
&b=b_{1}+b_{2}S^4,\\
&\mathcal{K}_i=\mathcal{K}_{i1}+\mathcal{K}_{i2}S^2,
\end{align}
where
\begin{align}
&a_1=-\frac{N_1}{2T_{c0}}\int\limits_{-\tilde\mu}^{1}dy\frac{\mbox{sech}^2(y/2\tilde{T}_{c0})}{\sqrt{y+\tilde\mu}},\\
&b_1=\frac{N_1}{4\hbar^2\omega_c^2}\int\limits_{-\tilde\mu}^{1}dy\frac{\mbox{sech}^2(y/2\tilde{T}_{c0})}{y^3\sqrt{y+\tilde\mu}}\nonumber\\
&\qquad\qquad\qquad\times\left[\sinh\left(\frac{y}{\tilde{T}_{c0}}\right)-\frac{y}{\tilde{T}_{c0}}\right],\\
&\mathcal{K}_{z1}=\hbar^2v_{1F}^2\frac{N_1}{8\hbar^2\omega_c^2}\int\limits_{-\tilde\mu}^{1}dy\frac{\sqrt{y+\tilde\mu}}{y^3}\mbox{sech}^2(y/2T_{c0})\nonumber\\
&\qquad\qquad\qquad\times\left[\sinh\left(\frac{y}{\tilde{T}_{c0}}\right)-\frac{y}{\tilde{T}_{c0}}\right]
\end{align}
and the coefficients for the terms from the deep band are widely known $a_2=N_2$, $b_2=N_2\frac{7\zeta(3)}{8\pi^2T_{c0}^2}$ and $\mathcal{K}_2=N_2\frac{7\zeta(3)}{8\pi^2T_{c0}^2}\frac{\hbar^2v_F^2}{2D}$, ($D=2,3$). In the expressions above, the only difference between a 2D and a 3D band in the deep band regime is the constant $N_2$, but as this constant can be hidden in the dimensionless coupling, we are able to perform a joint analysis for both cases. Note that $\mathcal{K}_{x1}=\mathcal{K}_{y1}=0$ for the q1D band and $\mathcal{K}_{x2}=0$ for the 2D variant of the stable band, due to very large effective electronic masses along these directions. Finally, the GL free energy for the composite systems q1D+2D or q1D+3D has actually a single-component order parameter, $\Psi({\bf r})$, because of the symmetry of the gap vector. In principle, fluctuations could enable a non-zero component also in the second eigenvector $\vec{\eta}_-$, but these fluctuations are non-critical and thus they can safely not be considered. 
Furthermore, the resulting free energy is effectively of a q1D+2D system and the corresponding Ginzburg-Levanyuk parameter (or Ginzburg number) can be expressed as
\begin{equation}
Gi=Gi^{2D}\frac{\frac{b_{1}}{b_{2}}+S^4}{4\pi |S|\left(\frac{a_{1}}{a_{2}}+S^2\right)\sqrt{\frac{\mathcal{K}_{x1}}{\mathcal{K}_{2}}+S^2}}
\end{equation}
where
\begin{equation}
Gi^{2D}=\frac{T_{c0}b_{2}}{4\pi a_{2}\mathcal{K}_{2}}.
\end{equation}
And similarly for the q1D+3D case the Ginzburg-Levanyuk parameter becomes
\begin{figure*}[t]
\centering
\includegraphics[width=0.3\linewidth]{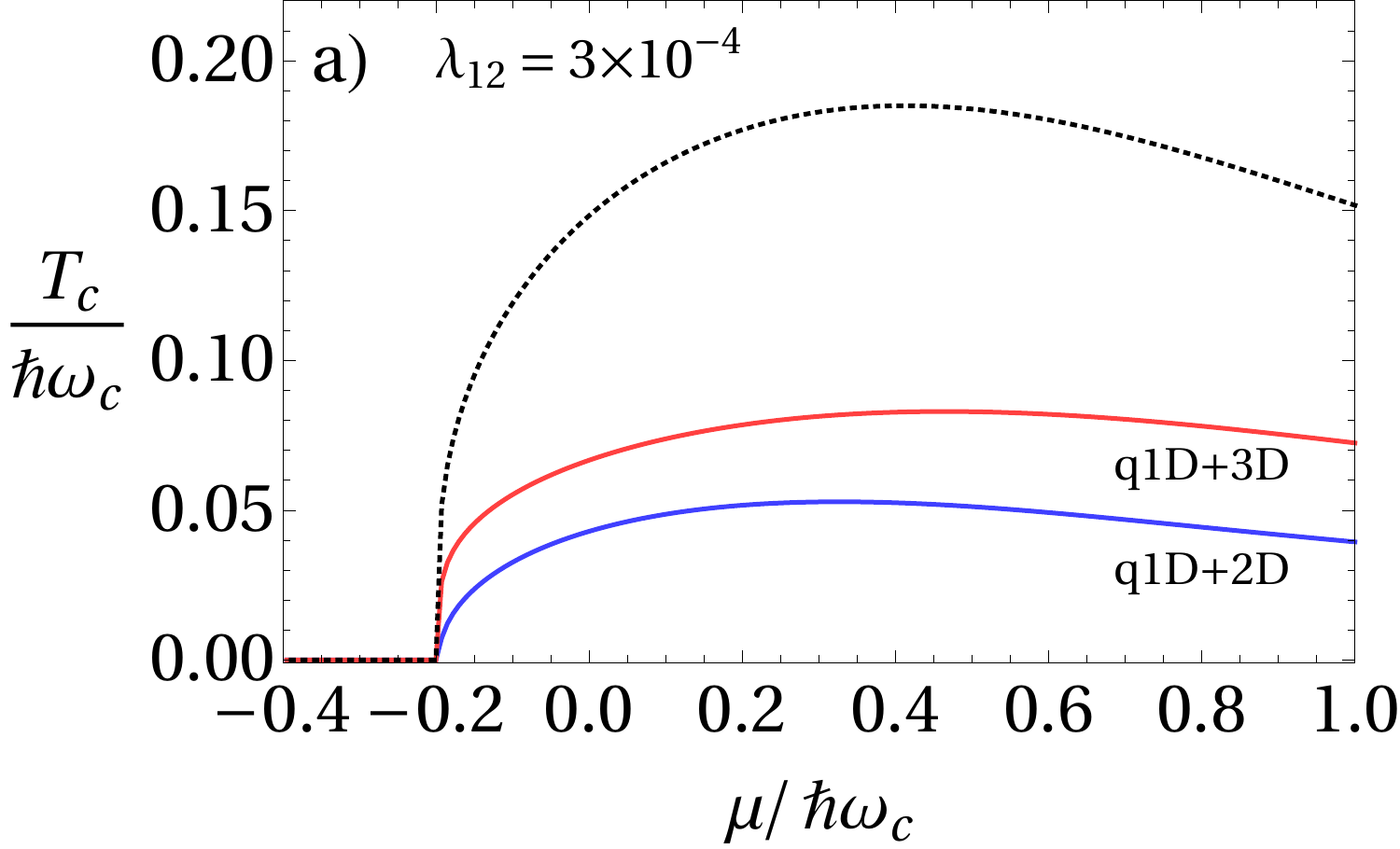}\hfill
\includegraphics[width=0.3\linewidth]{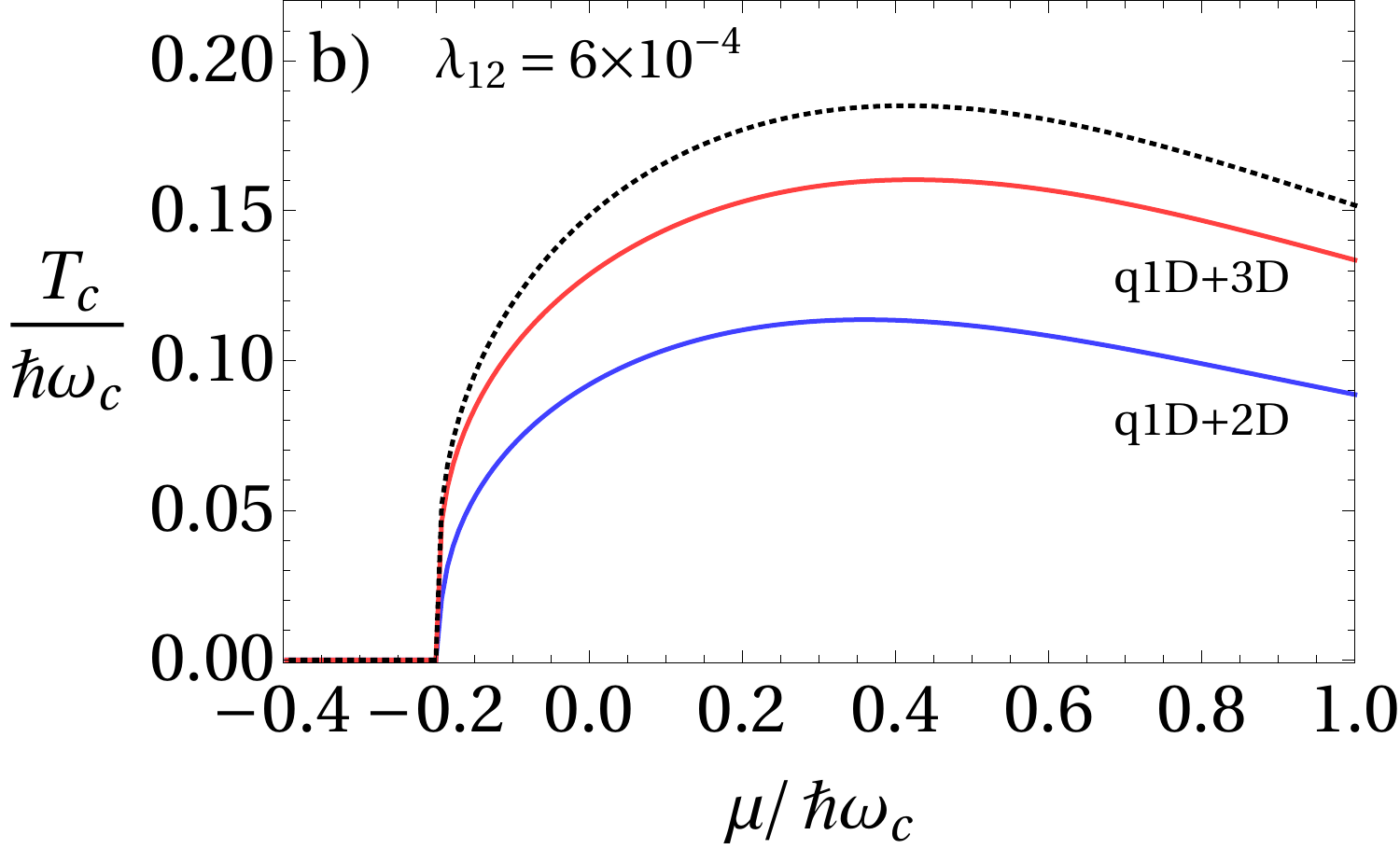}\hfill
\includegraphics[width=0.3\linewidth]{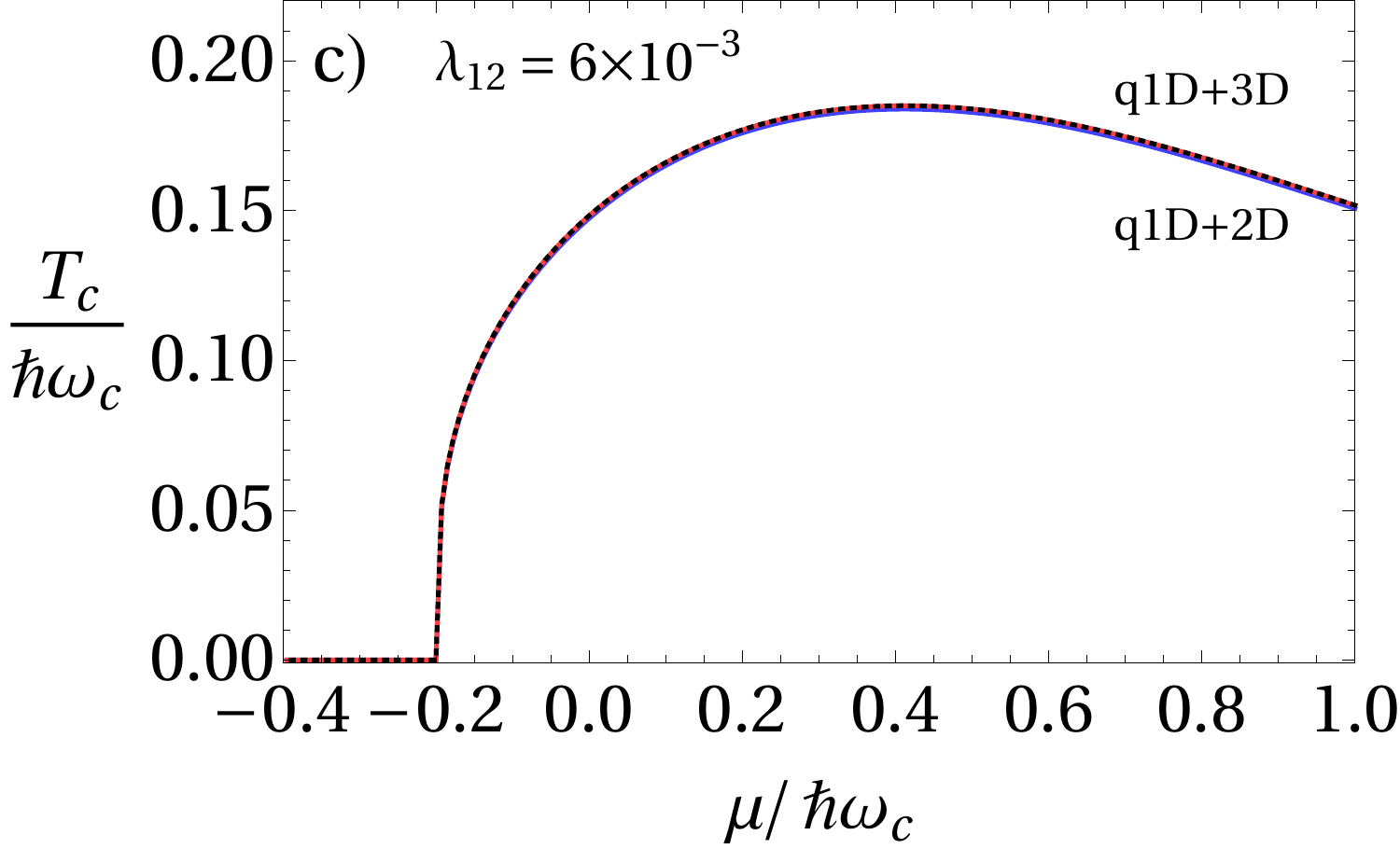}\\
\includegraphics[width=0.3\linewidth]{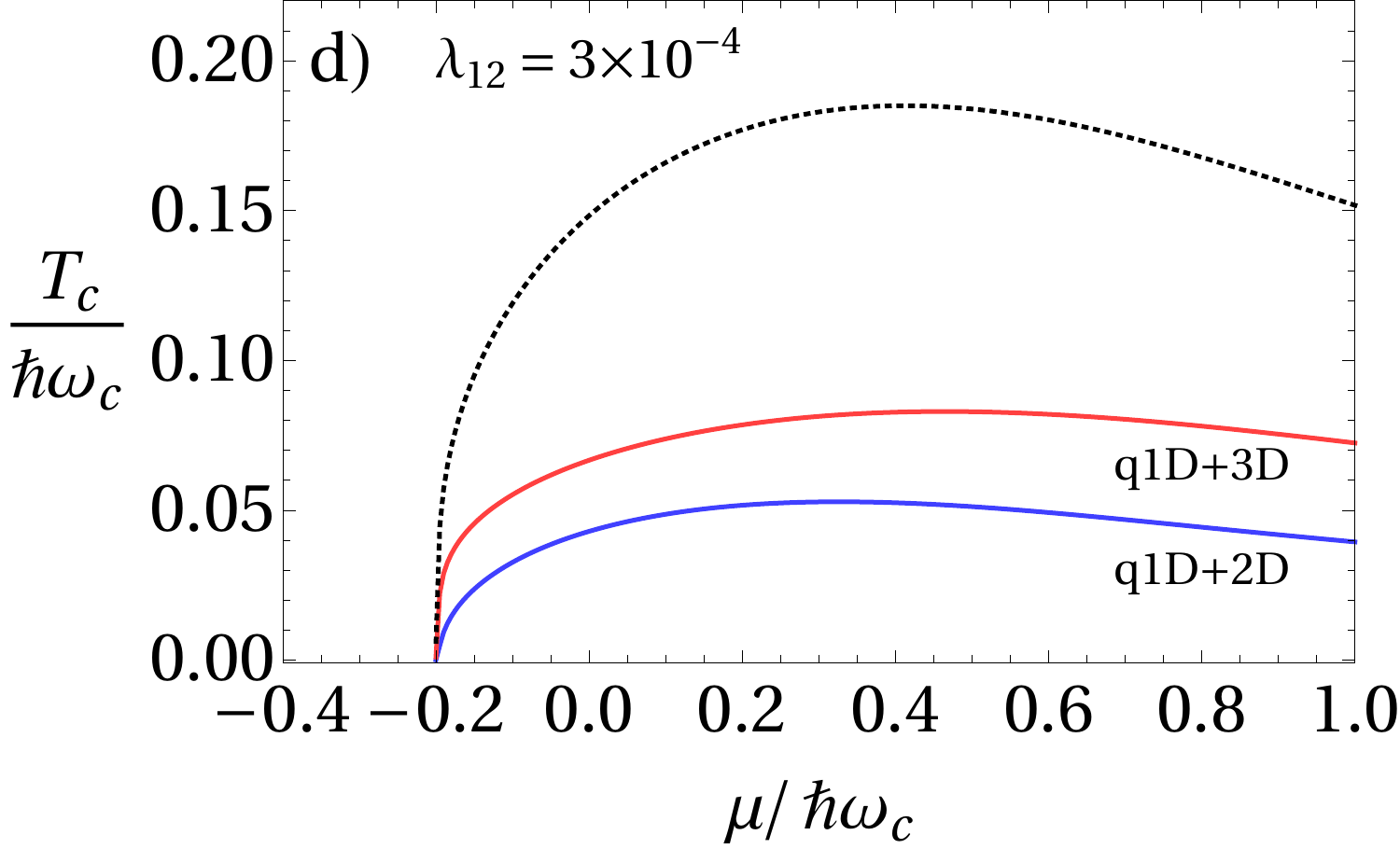}\hfill
\includegraphics[width=0.3\linewidth]{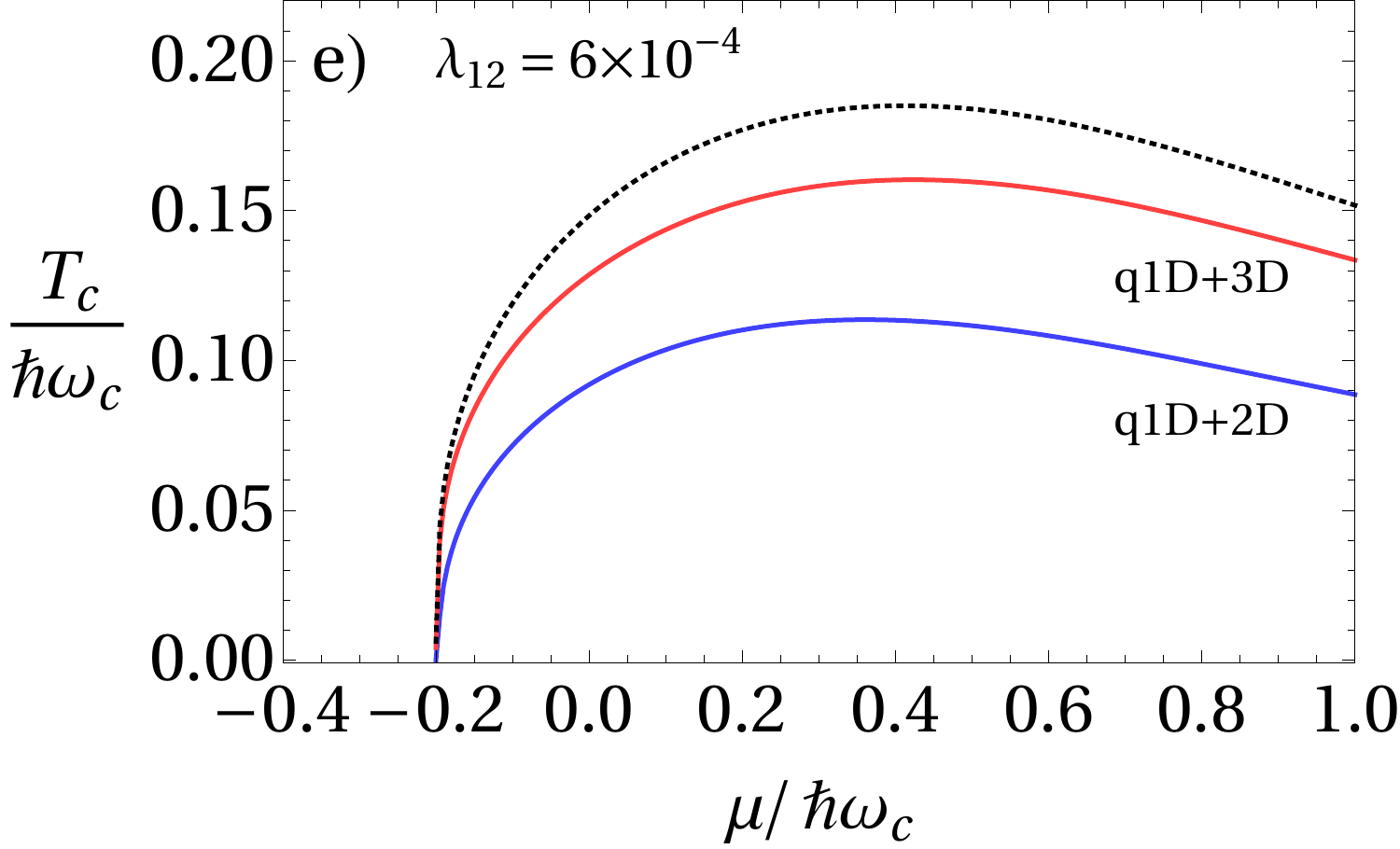}\hfill
\includegraphics[width=0.3\linewidth]{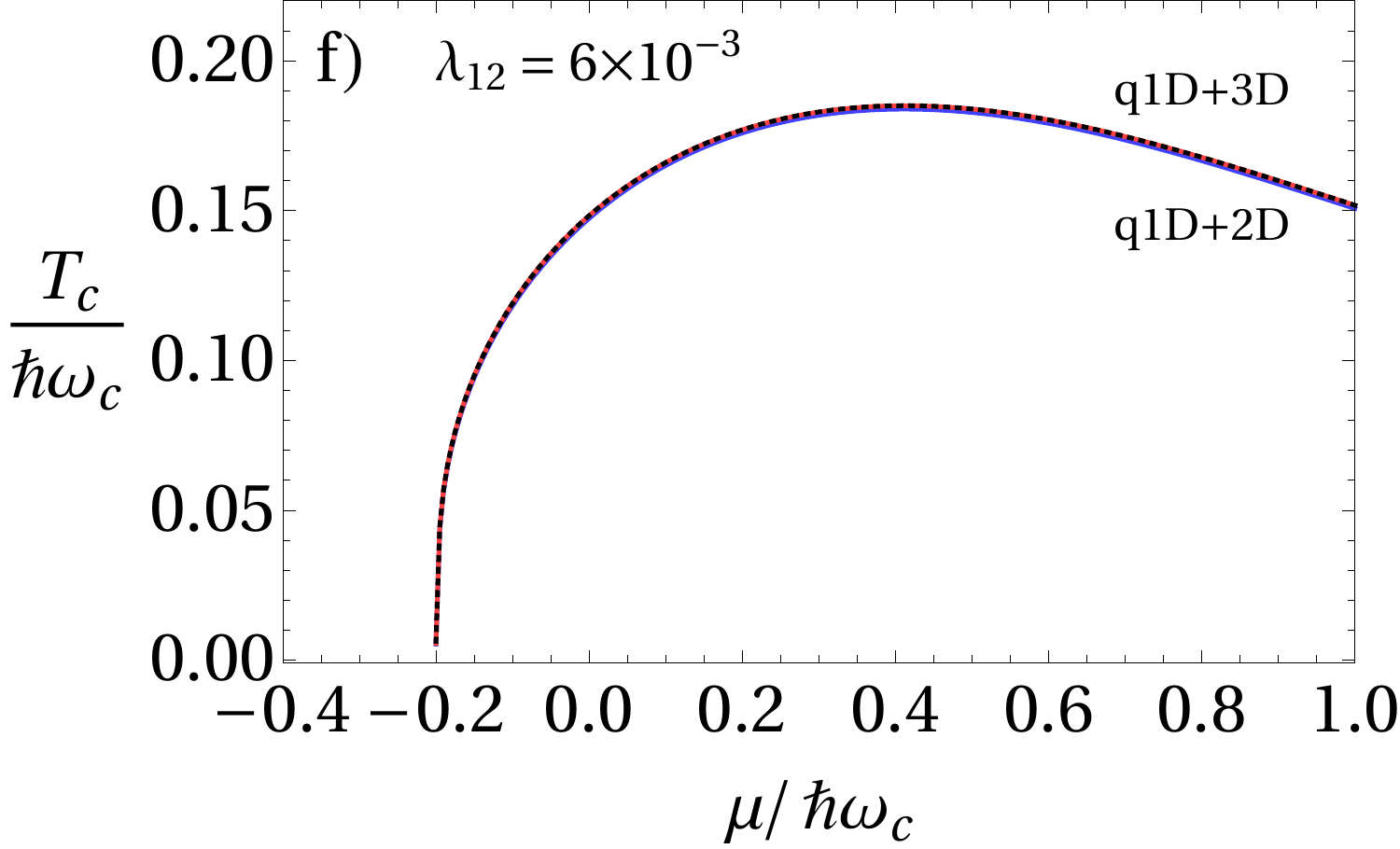}
\caption{\label{fig.Tc}
Renormalized critical temperatures due to fluctuations for q1D+2D (blue) and q1D+3D (red) systems for the parameters $\lambda_1=0.2$ and $\lambda_2=0.01$ upper row, plots a), b) and c) and $\lambda_2=0$ for the lower row, plots d), e) and f). The dotted black line is the mean-field critical temperature. We used the deep band typical Ginzburg number values for 2D and 3D deep bands $Gi^{(2D)}=10^{-5}$ and $Gi^{(3D)}=10^{-18}$, respectively.}
\end{figure*}
\begin{equation}
Gi=Gi^{3D}\frac{\left(\frac{b_{1}}{b_{2}}+S^4\right)^2}{\left(\frac{a_{1}}{a_{2}}+S^2\right)\left(\frac{\mathcal{K}_{x1}}{\mathcal{K}_{x2}}+S^2\right)S^4},
\end{equation}
where
\begin{equation}
Gi^{3D}=\frac{1}{32\pi^2}\frac{T_{c0}b_2^2}{a_2\mathcal{K}_2^3}.
\end{equation}
In the simple case where $v_{2T}/v_{1F}\to0$, i.e. the flat-band regime, we have
\begin{equation}
Gi=Gi^{2D}\frac{\frac{b_{1}}{b_{2}}+S^4}{4\pi S^2\left(\frac{a_{1}}{a_{2}}+S^2\right)}
\end{equation}
and
\begin{equation}
Gi=Gi^{3D}\frac{\left(\frac{b_1}{b_2}+S^4\right)^2}{\left(\frac{a_1}{a_2}+S^2\right)S^6}.
\end{equation}
for the q1D+2D and q1D+3D, respectively.
Finaly, the shift over the critical temperature can be written in terms of Gi as~\cite{Larkin,Salasnich2018,Saraiva2020}
\begin{equation}
\frac{T_{c0}-T_c}{T_c}=\frac{8}{\pi}Gi^{1/2},
\end{equation}
and
\begin{equation}
\frac{T_{c0}-T_c}{T_{c0}}=4Gi
\end{equation}
as the shift of the critical temperature due to the Berezinski-Kosterlitz-Thouless (BKT) transition for the 2D case.

\section{Results and Discussion}

The theoretical derivation shown in the previous section shows that, at the mean-field level, having a weaker 2D or a 3D deep bands should induce similar changes in the critical temperature. We show in Fig.~\ref{fig.tc0}
the mean-field critical temperature as a function of the chemical potential, both normalized by $\hbar\omega_c$. In both plots, we can see that there is a very strong increase of the critical temperature above $\mu/\hbar\omega_c\approx-0.2$, where the system goes through the so-called Lifshitz transition. In the first plot, a), the second band is much weaker but has nonzero coupling ($\lambda_2=0.01$), which means it would be superconducting even if uncoupled to the stronger band and this produces nonzero critical temperature in the region $\mu/\hbar\omega_c<0.2$, where there should not be pairs in the q1D band, once it is bellow the Lifshitz point. There are different values for the interband coupling from a very strong value, $\lambda_{12}=\lambda_1=0.2$, down to the limit of uncoupled bands $\lambda_{12}\to0$. For a non-passive second band ($\lambda_2>0$), the more coupled the bands, i.e. the higher is the value of $\lambda_{12}$, the higher is $T_{c0}$ and one can see that the curves have a maximum around $\mu/\hbar\omega_c\approx0.4$, after the point of divergence in the DOS for the q1D band where there is a sudden increase in $T_{c0}$. The dependence on $\lambda_{12}$ is very moderate and one can say that the main parameter in this system is the coupling in the shallow band, $\lambda_{11}$, because the deep band is taken with a small intraband coupling.

Finally, we demonstrate how the introduction of the second passive band induces the screening of fluctuations even in the extreme case of just a passive band. As can be seen in Figs.~\ref{fig.Tc} a) and d), for small interband couplings the fluctuations take over the superconducting phase and the renormalized critical temperatures can get much smaller than the mean-field solution. Now, the cases b), c) e and f), the stronger values of the interband coupling are enough to produce critical temperatures closer to the mean-field values. In the plot c), the value $\lambda_{12}=6\times10^{-3}$ is almost two orders of magnitude smaller than the coupling in the stronger band, $\lambda_1=0.2$. As can be seen, the difference between the upper and lower plots are negligible and therefore one concludes that the mechanism which we described is very robust and can work as a prototype for novel High-T$_{\mbox{c}}$ materials.

\section{Conclusion}
We showed a simple mechanism to stabilize fluctuations in a q1D superconductor where the q1D band is stronger and coupled to another weaker band with two or more dimensions.
This second band can be even just be a passive band where the Coopar pairs are formed in the stronger band and is exchanged to the weaker band.
The mean field solutions for the critical temperature in a single-band q1D system might be very high due to divergence of the DOS next to the bottom of the band, the Lifshitz point, but the shift of the critical temperature due to fluctuations is huge, making the superconducting state practically impossible in this case. In the case of a two-band system with a second band with a higher dimensional Fermi surface, this makes the material essentially higher dimensional which drastically reduces the renormalization of the critical temperature. Also, we showed how this mechanism is very robust once the second band can even be just a passive band, i.e. it would not be superconductor by itself. This mechanism captures both interesting qualities from the q1D, 2D and 3D systems: possible high critical temperatures next to the Lifshitz point and it shows little effect of fluctuations.

\appendix
\section{Calculation of the mean-field critical temperature, $\mathbf{T_{c0}}$, and the GL coefficients}\label{SecApp}

Following the Green function formalism developed in Ref's.~\onlinecite{Gorkov1958,AGD1965}, the Hamiltonian given in Eq.~(\ref{eq.H}) allows us to construct Dyson-like equations for the anomalous averages in terms of the normal-state temperature Green functions $\mathcal{G}_{\nu\omega}^{(0)}(\bf x,\bf y)$ and $\bar{\mathcal{G}}_{\nu\omega}^{(0)}(\bf x,\bf y)$:
\begin{align}\label{ap.gapeq}
&R_\nu[\Delta_\nu]=\int d^3{\bf y} K_{\nu a}({\bf x},{\bf y})\Delta_\nu({\bf y})\nonumber\\
&\qquad\qquad+\int \prod_{l=1}^3d^3{\bf y}_lK_{\nu b}({\bf x},{\bf y}_1,{\bf y}_2,{\bf y}_3)\nonumber\\
&\qquad\qquad\qquad\qquad\times\Delta_\nu({\bf y}_1)\Delta_\nu^\ast({\bf y}_2)\Delta_\nu({\bf y}_3),
\end{align}
where the kernels are given by
\begin{align}
K_{\nu a}({\bf x},{\bf y}) = -gT\sum_\omega\mathcal{G}_{\nu\omega}^{(0)}({\bf x},{\bf y})\bar{\mathcal{G}}_{\nu\omega}^{(0)}({\bf y},{\bf x})
\end{align}
and
\begin{align}
&K_{\nu b}({\bf x},{\bf y}_1,{\bf y}_2,{\bf y}_3)=-gT\sum_\omega \mathcal{G}_{\nu\omega}^{(0)}({\bf x},{\bf y}_1)\nonumber\\
&\qquad\qquad\qquad\times\bar{\mathcal{G}}_{\nu\omega}^{(0)}({\bf y}_1,{\bf y}_2)\mathcal{G}_{\nu\omega}^{(0)}({\bf y}_2,{\bf y}_3)\bar{\mathcal{G}}_{\nu\omega}^{(0)}({\bf y}_3,{\bf x}).
\end{align}
The normal-state temperature Green functions are defined in terms of the band-dependent single electron energies, $\xi_k^{(\nu)}$, as
\begin{equation}
\mathcal{G}_{\nu\omega}^{(0)}({\bf x},{\bf y})=\int \frac{d^3{\bf k}}{(2\pi)^3} \frac{e^{-i{\bf k}({\bf x}-{\bf y})}}{i\hbar\omega-\xi_k^{(\nu)}}
\end{equation}
and $\bar{\mathcal{G}}_{\nu\omega}^{(0)}({\bf x},{\bf y})=-\mathcal{G}^{(0)}_{\nu,-\omega}({\bf y},{\bf x})$. The integral kernels involve, as usual, the summation over the fermionic Matsubara frequencies $\omega_n=\pi T(2n+1)/\hbar$ (here the Boltzmann constant $k_B$ is set to $1$).

The effective dimensions of the Fermi sheets are considered in the regime when the dispersion relation has very large effective electronic masses in some directions, say, $m_y,m_z\gg m_x$ (for the q1D case) or $m_z\gg m_x,m_y$ (for the 2D case). Then the related single-particle energy becomes
\begin{equation}\label{ap.disprel}
\xi_k=\sum_{i=1}^3\frac{\hbar^2k_i^2}{2m_i}-\mu\approx\left\{\begin{array}{lc}
\frac{\hbar^2k_x^2}{2m_x}-\mu&\mbox{\ (1D)}\\
\frac{\hbar^2k_x^2}{2m_x}+\frac{\hbar^2k_y^2}{2m_y}-\mu& \mbox{\ (2D)}\\
\frac{\hbar^2k^2}{2m}-\mu& \mbox{\ (3D)}
\end{array}\right.
\end{equation}
In the so called deep band regime, one can shift the bottom of the 2D or 3D bands by the constant $\varepsilon_0$, as stated in Sec.~\ref{secII}.

Let us begin with the linearized version of Eq.~(\ref{ap.gapeq}) for a system with a q1D band and a second band with higher number of dimensions $D=2$ or $3$ and contract with the matrix $g_{\nu\nu'}$:
\begin{align}
&\Delta_{\nu'}=\sum_{\nu=1,2} g_{\nu\nu'}R_{\nu}=-T_{c0}\sum_{\nu=1,2}g_{\nu\nu'}\Delta_\nu\nonumber\\
&\quad\times\sum_\omega\int d{\bf z}\frac{d{\bf k}d{\bf k'}}{(2\pi)^6}\frac{\exp[-i({\bf k}-{\bf k}')\cdot{\bf z}]}{\left[i\hbar\omega-\xi_k^{(\nu)}\right]\left[i\hbar\omega+\xi_{k'}^{(\nu)}\right]}\label{ap.mateq}
\end{align}
where ${\bf z}={\bf x}-{\bf y}$. The two coefficients of $\Delta_\nu$ in Eq.~(\ref{ap.mateq}) can be rewritten as
\begin{equation}
-g_{1\nu'}T_{c0}\sum_\omega\sigma^{(z)}\int\frac{dk}{2\pi}\frac{1}{(i\hbar\omega-\xi_k^{(1)})(i\hbar\omega+\xi_k^{(1)})}.
\end{equation}
for a 2D band.
\begin{equation}
-g_{1\nu'}T_{c0}\sum_\omega\sigma^{(yz)}\int\frac{dk_x}{2\pi}\frac{1}{(i\hbar\omega-\xi_k)(i\hbar\omega+\xi_k^{(1)})},
\end{equation}
Here we defined the auxiliary parameters
\begin{align}
\sigma^{(z)}&=\int\frac{dk_z}{2\pi}=\frac{k_F^{(z)}}{\pi},\\
\sigma^{(yz)}&=\int\frac{dk_y}{2\pi}\frac{dk_z}{2\pi}=\int\frac{dk}{2\pi}=\frac{k_F^{(yz)}}{2\pi}.
\end{align}
Furthermore, considering rotation symmetry ($m_x=m_y=m$), we have:
\begin{align}
d\xi&=\frac{\hbar^2k_x}{m_x}dk_x\Rightarrow dk_x\to\sqrt{\frac{m_x}{2\hbar^2}}\ \frac{d\xi}{\sqrt{\mu+\xi}},\\
d\xi&=\frac{\hbar^2k}{m}dk\Rightarrow \frac{dk_xdk_y}{(2\pi)^2}=\frac{kdk}{2\pi}\to \frac{m}{\hbar^2}d\xi.
\end{align}
This means that the integrals can be written as
\begin{align}
&T_{c0}g\sigma^{(yz)}\sqrt{\frac{m_x}{8\pi^2\hbar^2}}\sum_\omega\int_{-\mu}^{\hbar\omega_c}d\xi \frac{(\mu+\xi)^{-1/2}}{\hbar^2\omega^2+\xi^2}=1\\
&T_{c0}g\sigma^{(z)}\frac{m}{\hbar^2}\sum_\omega\int_{-\mu}^{\hbar\omega_c} d\xi \frac{1}{\hbar^2\omega^2+\xi^2}=1
\end{align}
and that the DOS for the q1D band becomes (by introducing the excitation energy independent from the chemical potential $E=\xi+\mu$):
\begin{equation}
N_{1d}(E)=\sigma^{(yz)}\sqrt{\frac{m_x}{8\pi^2\hbar^2E}}
\end{equation}
while the DOS for the 2D system is a constant
\begin{equation}
N_{2d}=\sigma^{(z)}\frac{m}{\hbar^2}.
\end{equation}
The summation over the Matsubara frequencies is known
\begin{equation}
\sum_\omega\frac{1}{\hbar^2\omega^2+\xi^2}=\frac{\tanh(\xi/2T)}{2T\xi}
\end{equation}
and then
\begin{align}\label{ap.summed}
&g\sigma^{(yz)}\sqrt{\frac{m_x}{32\pi^2\hbar^2}}\int\limits_{0}^{\hbar\omega_c+\mu} dE\ \frac{\tanh[(E-\mu)/2T_{c0}]}{(E-\mu)E^{1/2}}=1,\\
&g\sigma^{(z)}\frac{m}{2\hbar^2}\int\limits_{\varepsilon_0}^{\hbar\omega_c+\mu}dE\frac{\tanh[(E-\mu)/2T_{c0}]}{E-\mu}=1,
\end{align}
where it was introduced the cutoff energy $\hbar\omega_c$. Although the integral appearing in Eq.~(\ref{ap.summed}) is not divergent, we introduce a physical cutoff as was done in the 3D case.
It is natural to introduce the dimensionless couplings
\begin{align}
&\lambda_{1d}=g_{11}\sigma^{(yz)}\sqrt{\frac{m_x}{32\pi^2\hbar^3\omega_c}}\\
&\lambda_{2d}=g_{22}\sigma^{(z)}\frac{m}{2\hbar^2}
\end{align}
and, writing the relevant quantities in units of $\hbar\omega_c$, the equations for $T_{c0}$ becomes:
\begin{align}
&\lambda_{1d}\int_{0}^{1+\tilde\mu} dx\ \frac{\tanh[(x-\tilde\mu)/2T_{c0}]}{(x-\tilde\mu)x^{1/2}}=1\\
&\lambda_{2d}\int_0^{1+\tilde\mu} dx\frac{\tanh[(x-\tilde\mu)/2\tilde{T}_{c0}]}{x-\tilde\mu}=1.
\end{align}
In the deep band regime, the equation for the 2D band becomes:
\begin{align}
&\lambda_{2d}\int\limits_{0}^{\infty}dx\frac{\tanh(x/2\tilde T_{c0})}{x}=1\Rightarrow\nonumber\\
&\tilde T_{c0}=\frac{2e^{\Gamma}}{\pi}\exp\left(-1/\lambda_{2d}\right)
\end{align}

In order to include the effect of fluctuations of the gap, we consider the deviation from the critical temperature, $\tau=1-T/T_c$, and we will consider the first gradient terms of the Taylor expansion of the gap in the linear term
\begin{align}
&\int d^3zK_a({\bf z})\Delta({\bf z})\nonumber\\
&\qquad\approx\int d^3zK_a({\bf z})\left[\Delta({\bf x})+\frac{({\bf z}\cdot\vec{\nabla})^2}{2}\Delta({\bf x})\right].
\end{align}
We can obtain the first GL coefficients, $a_{1d}$ and $a_{2d}$, by deriving with respect to $\tau$ the hyperbolic tangent:
\begin{align}\label{aq1D}
a_{1d}&=\frac{N_{1d}\sqrt{\hbar\omega_c}}{2T_{c0}}\int\limits_{0}^{\hbar\omega_c+\mu}\mbox{d}E\frac{\text{sech}^2\left[(E-\mu)/2T_{c0}\right]}{E^{1/2}}\\
a_{2d}&=\frac{N_{2d}}{2T_{c0}}\int\limits_0^{\hbar\omega_c+\mu}dE\ \text{sech}^2\left[(E-\mu)/2T_{c0}\right]\nonumber\\
&=N_{2d}\left[1+\tanh(\mu/2T_{c0})\right]
\end{align}

The second term is composed by integrals such as
\begin{align}
I_{\mathcal{K}}^{(i,j)}&=-T\sum_\omega\int\mbox{d}^3z\frac{\mbox{d}^3k}{(2\pi)^3}\frac{\mbox{d}^3k'}{(2\pi)^3}\frac{z_iz_j}{2}\times\nonumber\\
&\qquad\times\frac{\exp[-i({\bf k}-{\bf k}')\cdot{\bf z}]}{(i\hbar\omega-\xi_k)(i\hbar\omega+\xi_{k'})},
\end{align}
but due to the symmetry of the integrands, $I_{a2}^{(i,j)}=0$ for $i\neq j$.
The terms $z_i\mathcal{G}_\omega^{(0)}({\bf z})$ can be replaced by the derivative with respect to $k_i$ in the $k-$space and, again, the volume integration over ${\bf z}$ produces $\delta({\bf k}'-{\bf k})$ and then
\begin{align}
I_{\mathcal{K}}^{(i,i)}&=-\frac{T_{c0}}{2}\sum_\omega\int \frac{\mbox{d}^3k}{(2\pi)^3}\left(\partial_{k_i}\frac{1}{i\hbar\omega-\xi_k}\right)\times\nonumber\\
&\qquad\times\left(\partial_{k_i}\frac{1}{i\hbar\omega+\xi_k}\right).
\end{align}
It is trivial that $I_{a2}^{(i,i)}=0$ for $i=y,z$ in the q1D case and for $i=z$ in the 2D case.
Finally, in the q1D case for $i=x$, we have
\begin{align}
I_{\mathcal{K}}^{(x,x)}&=-\frac{T_{c0}}{2}\sum_\omega\int \frac{\mbox{d}^3k}{(2\pi)^3}\frac{-\left(\frac{\hbar^2k_x}{m_x}\right)^2}{(\hbar^2\omega^2+\xi_k^2)^2}\\
&=\frac{\hbar^2}{m_x}T_{c0}\sum_\omega\int \frac{\mbox{d}^3k}{(2\pi)^3} \frac{\xi_k+\mu}{(\hbar^2\omega^2+\xi_k^2)^2}\\
&=\frac{\hbar^2}{m_x}T_{c0}\sigma^{(xy)}\sum_\omega\int \frac{\mbox{d}k_x}{2\pi} \frac{\xi_k+\mu}{(\hbar^2\omega^2+\xi_k^2)^2}
\end{align}
and for $i=x,y$ in the 2D case, we have (again considering $m_x=m_y=m$)
\begin{align}
I_{\mathcal{K}}^{(i,i)}&=-\frac{T_{c0}}{2}\sum_\omega\int \frac{\mbox{d}^3k}{(2\pi)^3}\frac{-\left(\frac{\hbar^2k_i}{m}\right)^2}{(\hbar^2\omega^2+\xi_k^2)^2}\\
&=\frac{\hbar^2}{m}T_{c0}\sum_\omega\int \frac{\mbox{d}^3k}{(2\pi)^3} \frac{(\xi_k+\mu)/2}{(\hbar^2\omega^2+\xi_k^2)^2}\\
&=\frac{\hbar^2}{2m}T_{c0}\sigma^{(z)}\sum_\omega\int \frac{k\mbox{d}k}{2\pi} \frac{\xi_k+\mu}{(\hbar^2\omega^2+\xi_k^2)^2}
\end{align}
and here we use the tabled infinite summation over Matsubara frequencies
\begin{equation}\label{sum2}
\sum_\omega\frac{1}{(\hbar^2\omega^2+\xi_k^2)^2}=
\frac{\left[T \sinh\left(\xi_k/T\right)-\xi_k\right] \text{sech}^2\left(\frac{\xi_k}{2 T}\right)}{8 \xi_k^3 T^2}
\end{equation}
Then:
\begin{align}
\mathcal{K}_{1d}^{(x)}&=\frac{\hbar^2}{m_x}T_{c0}\sigma^{(yz)}\sqrt{\frac{m_x}{8\pi^2\hbar^2}}\int\limits_0^{\hbar\omega_c+\mu} \mbox{d}EE^{1/2}\nonumber\\
&\qquad\times\frac{\text{sech}^2\left[(E-\mu)/2T_{c0}\right]}{8(E-\mu)^3T_{c0}}\nonumber\\
&\qquad\times\left[\sinh\left(\frac{E-\mu}{T_{c0}}\right)-\frac{E-\mu}{T_{c0}}\right]\\
&=\frac{\hbar^2}{m_x}\frac{N_{1d}\sqrt{\hbar\omega_c}}{4}\int\limits_0^{\hbar\omega_c+\mu} \mbox{d}EE^{1/2}\frac{\text{sech}^2\left[(E-\mu)/2T_{c0}\right]}{(E-\mu)^3}\nonumber\\
&\qquad\times\left[\sinh\left(\frac{E-\mu}{T_{c0}}\right)-\frac{E-\mu}{T_{c0}}\right].
\label{k1d}
\end{align}
For the 2D case, one has the shallow band version:
\begin{align}
\mathcal{K}_{2d}^{(i)}&=\frac{1}{4\pi }T_{c0}\sigma^{(z)}\int\limits_0^{\hbar\omega_c+\mu} \mbox{d}E\frac{\text{sech}^2\left[(E-\mu)/2T_{c0}\right]}{8(E-\mu)^3T_{c0}}\times\nonumber\\
&\qquad\times\left[\sinh\left(\frac{E-\mu}{T_{c0}}\right)-\frac{E-\mu}{T_{c0}}\right]\\
&=\frac{\hbar^2}{m}\frac{N_{2d}}{32\pi}\int\limits_0^{\hbar\omega_c+\mu} \mbox{d}E\frac{\text{sech}^2\left[(E-\mu)/2T_{c0}\right]}{(E-\mu)^3}\nonumber\\
&\qquad\times\left[\sinh\left(\frac{E-\mu}{T_{c0}}\right)-\frac{E-\mu}{T_{c0}}\right]
\label{k1d}
\end{align}
and the stiffness of the gap parameter along the other orthogonal directions is zero.
In the deep band regime, 
\begin{align}
\mathcal{K}_{2d}^{(i)}&=\frac{\hbar^2}{2}\frac{\mu}{m}T_{c0}\sigma^{(z)}\mu\frac{m}{\hbar^2}\sum_\omega\int\limits_{-\infty}^{\infty} d\xi\frac{1}{(\hbar^2\omega^2+\xi_k^2)^2}\nonumber\\
&=\frac{\hbar^2v_F^2}{4}T_{c0}N_{2d}\sum_\omega\frac{1}{|\hbar\omega|^3}2\int\limits_0^\infty dx\frac{1}{(1+x^2)^2}\nonumber\\
&=\frac{\hbar^2v_F^2}{2}N_{2d}\frac{\pi}{2}\frac{1}{\pi^3T_{c0}^2}\sum\limits_{n=0}^\infty\frac{1}{(2n+1)^3}\nonumber\\
&=\hbar^2v_F^2\frac{7\zeta(3)N_{2d}}{32\pi^2T_{c0}^2}
\end{align}

The last term is given by the cubic contribution from Eq.~(\ref{ap.gapeq}).
It is enough to consider only the zero-order contribution of the gap in the Taylor expansion on the coordinates (i.e. it becomes independent of the gap) and thus the integral becomes
\begin{align}
b&=\int\prod_{l=1}^3d^3y_lK_b({\bf x},{\bf y}_1,{\bf y}_2,{\bf y}_3)\\
&=-T\sum_\omega\int\frac{\mbox{d}^3k}{(2\pi)^3}\frac{1}{(i\hbar\omega-\xi_k)^2(i\hbar\omega+\xi_k)^2},
\end{align}
where we used the convolution theorem to find the Fourier transform of the product of unperturbed Green functions.
Next we apply the summation given by Eq.~(\ref{sum2}) and the final expression for the coefficient $b$ for the q1D case becomes
\begin{align}
b_{1d}&=T_{c0}\sigma^{(xy)}\sqrt{\frac{m_x}{32\pi^2\hbar^2}}\int\limits_0^{\hbar\omega_D+\mu}\mbox{d}E\frac{\text{sech}^2\left[(E-\mu)/2T_{c0}\right]}{8T_{c0}^2E^{1/2}(E-\mu)^3}\times\nonumber\\
&\qquad\times\left[T_{c0}\sinh\left(\frac{E-\mu}{T_{c0}}\right)-(E-\mu)\right]\\
&=\frac{N_{1d}\sqrt{\hbar\omega_c}}{8}\int\limits_0^{\hbar\omega_c+\mu}\mbox{d}E\frac{\text{sech}^2\left[(E-\mu)/2T_{c0}\right]}{E^{1/2}(E-\mu)^3}\times\nonumber\\
&\qquad\times\left[\sinh\left(\frac{E-\mu}{T_{c0}}\right)-\frac{E-\mu}{T_{c0}}\right]
\label{bq1D}
\end{align}
and for the 2D case, in the shallow band regime it becomes
\begin{align}
b_{2d}&=T_{c0}\sigma^{(z)}\frac{m}{\hbar^2}\int\limits_0^{\hbar\omega_c+\mu}\mbox{d}E\frac{\mbox{sech}[(E-\mu)/2T_{c0}]}{8T_{c0}^2(E-\mu)^3}\times\nonumber\\
&\qquad\times\left[T_{c0}\sinh\left(\frac{E-\mu}{T_{c0}}\right)-(E-\mu)\right]\\
&=\frac{N_{2d}}{4}\int\limits_0^{\hbar\omega_c+\mu}\mbox{d}E\frac{\mbox{sech}[(E-\mu)/2T_{c0}]}{(E-\mu)^3}\times\nonumber\\
&\qquad\times\left[\sinh\left(\frac{E-\mu}{T_{c0}}\right)-\frac{E-\mu}{T_{c0}}\right]
\end{align}
and in the deep band regime it becomes
\begin{align}
b_{2d}&=T_{c0}\sigma^{(z)}\frac{m}{\hbar^2}\sum_\omega\frac{1}{|\hbar\omega|^3}2\int\limits_0^\infty dx\frac{1}{(1+x^2)^2}\nonumber\\
&=T_{c0}N_{2d}\frac{\pi}{2}\sum_\omega\frac{1}{|\hbar\omega|^3}\nonumber\\
&=\frac{N_{2d}}{\pi^2T_{c0}^2}\sum\limits_{n=0}^\infty\frac{1}{(1+2n)^3}\nonumber\\
&=\frac{7\zeta(3)N_{2d}}{8\pi^2T_{c0}^2}
\end{align}

One can simplify Eq's.~(\ref{aq1D}),~(\ref{k1d}) and~(\ref{bq1D}) by expressing all energies in units of $T_{c0}$ after noticing that the upper limit
\begin{equation}
\frac{\hbar\omega_c+\mu}{T_{c0}}\gtrsim10\to\infty
\end{equation}
for all the range of parameters we have used and that all the terms integrated in these equations are fast-decaying because of the term $\mbox{sech}^2[(E-\mu)/2T_{c0}]$.
These expressions are
\begin{align}
&a_{1d}=\frac{N_{1d}\sqrt{\hbar\omega}}{2T_{c0}^{1/2}}\int\limits_0^\infty \mbox{d}x\frac{\mbox{sech}^2[(x-\mu')/2]}{x^{1/2}},\\
&b_{1d}=\frac{N_{1d}\sqrt{\hbar\omega_c}}{8T_{c0}^{5/2}}\int\limits_0^\infty\mbox{d}x\frac{\mbox{sech}^2[(x-\mu')/2]}{x^{1/2}(x-\mu')^2}\left[\frac{\mbox{sinh}(x-\mu')}{x-\mu'}-1\right],\\
&\mathcal{K}_{1d}=\frac{\hbar^2}{m_x}\frac{N_s\sqrt{\hbar\omega_c}}{4T_{c0}^{3/2}}\int\limits_0^\infty\mbox{d}x
\frac{x^{1/2}\mbox{sech}^2[(x-\mu')/2]}{(x-\mu')^2}\times\nonumber\\
&\qquad\qquad\qquad\times\left[\frac{\sinh(x-\mu')}{x-\mu'}-1\right],
\end{align}
where $\mu'=\mu/T_{c0}$. The only reason for the appearance of $\hbar\omega_c$ in the equations is to maintain the term $N_{1d}$ constant and with units of DOS.
So in the algorithm to calculate the GL parameters, one must first calculate $T_{c0}/\hbar\omega_c$ and then calculate the GL parameters.
Note that, $\hbar\omega_c$ is not present in the expression for $Gi$.

\bibliographystyle{unsrt}
\bibliography{GeneralBibliography}

\begin{thebibliography}{10}

\bibitem{Nagamatsu2001}
J.~Nagamatsu, N.~Nakagawa, T.~Murakana, Y.~Zenitani, and J.~Akumitsu.
\newblock Superconductivity at 39k in magnesium diboride.
\newblock {\em Nature}, 410, 2001.

\bibitem{Larbalestier2001}
D.~C. Larbalestier, L.~D. Cooley, M.~O. Rikel, A.~A. Polyanskii, J.~Jiang,
  S.~Patnaik, X.~Y. Cai, D.~M. Feldmann, A.~Gurevich, A.~A. Squitieri, M.~T.
  Naus, C.~B. Eom, E.~E. Hellstrom, R.~J. Cava, K.~A. Regan, N.~Rogado, M.~A.
  Hayward, T.~He, J.~S. Slusky, P.~Khalifah, K.~Inumaru, and M.~Haas.
\newblock Strongly linked current flow in polycrystalline forms of the
  superconductor {MgB$_2$}.
\newblock {\em Nature}, 410(6825):186--189, Mar 2001.

\bibitem{Orlova2013}
N.~V. Orlova, A.~A. Shanenko, M.~V. Milo\ifmmode \check{s}\else
  \v{s}\fi{}evi\ifmmode~\acute{c}\else \'{c}\fi{}, F.~M. Peeters, A.~V. Vagov,
  and V.~M. Axt.
\newblock {G}inzburg-{L}andau theory for multiband superconductors: Microscopic
  derivation.
\newblock {\em Phys. Rev. B}, 87:134510, Apr 2013.

\bibitem{Milosevic2015}
Milorad~V Milo{\v{s}}evi{\'{c}} and Andrea Perali.
\newblock Emergent phenomena in multicomponent superconductivity: an
  introduction to the focus issue.
\newblock {\em Superconductor Science and Technology}, 28(6):060201, apr 2015.

\bibitem{Huang20}
Wen-Min Huang and Hsiu-Hau Lin.
\newblock Pairing mechanism in multiband superconductors.
\newblock {\em Scientific Reports}, 10(1):7439, May 2020.

\bibitem{Salasnich2018}
L.~Salasnich, A.~A. Shanenko, A.~Vagov, J.~Albino Aguiar, and A.~Perali.
\newblock Screening of pair fluctuations in superconductors with coupled
  shallow and deep bands: a route to higher temperature superconductivity.
\newblock arXiv:1810.03321, 2018.

\bibitem{Saraiva2020}
T.~T. Saraiva, P.~J.~F. Cavalcanti, A.~Vagov, A.~S. Vasenko, A.~Perali,
  L.~Dell'Anna, and A.~A. Shanenko.
\newblock Multiband material with a quasi-1d band as a robust high-temperature
  superconductor.
\newblock {\em Phys. Rev. Lett.}, 125:217003, Nov 2020.

\bibitem{Bao2015}
Jin-Ke Bao, Ji-Yong Liu, Cong-Wei Ma, Zhi-Hao Meng, Zhang-Tu Tang, Yun-Lei Sun,
  Hui-Fei Zhai, Hao Jiang, Hua Bai, Chun-Mu Feng, Zhu-An Xu, and Guang-Han Cao.
\newblock Superconductivity in quasi-one-dimensional
  {${\mathrm{K}}_{2}{\mathrm{Cr}}_{3}{\mathrm{As}}_{3}$} with significant
  electron correlations.
\newblock {\em Phys. Rev. X}, 5:011013, Feb 2015.

\bibitem{Tang2015A}
Zhang-Tu Tang, Jin-Ke Bao, Yi~Liu, Yun-Lei Sun, Abduweli Ablimit, Hui-Fei Zhai,
  Hao Jiang, Chun-Mu Feng, Zhu-An Xu, and Guang-Han Cao.
\newblock Unconventional superconductivity in quasi-one-dimensional
  {${\mathrm{Rb}}_{2}{\mathrm{Cr}}_{3}{\mathrm{As}}_{3}$}.
\newblock {\em Phys. Rev. B}, 91:020506, Jan 2015.

\bibitem{Wu2019}
Si-Qi Wu, Chao Cao, and Guang-Han Cao.
\newblock {Lifshitz transition and nontrivial H-doping effect in the Cr-based
  superconductor ${\mathrm{KCr}}_{3}{\mathrm{As}}_{3}{\mathrm{H}}_{x}$}.
\newblock {\em Phys. Rev. B}, 100:155108, Oct 2019.

\bibitem{Wang2017a}
Ren-Shu Wang, Yun Gao, Zhong-Bing Huang, and Xiao-Jia Chen.
\newblock Superconductivity in p-terphenyl.
\newblock {\em arxiv.org}, 2017.

\bibitem{Wang2017b}
Ren-Shu Wang, Yun Gao, Zhong-Bing Huang, and Xiao-Jia Chen.
\newblock Superconductivity at 43 k in a single c-c bond linked terphenyl.
\newblock {\em arxiv.org}, 2017.

\bibitem{Wang2017c}
Ren-Shu Wang, Yun Gao, Zhong-Bing Huang, and Xiao-Jia Chen.
\newblock Superconductivity above 120 kelvin in a chain link molecule.
\newblock {\em arxiv.org}, 2017.

\bibitem{Suhl1959}
H.~Suhl, B.~T. Matthias, and L.~R. Walker.
\newblock {B}ardeen-{C}ooper-{S}chrieffer theory of superconductivity in the
  case of overlapping bands.
\newblock {\em Phys. Rev. Lett.}, 3:552--554, Dec 1959.

\bibitem{Moskalenko1959}
V.~A. Moskalenko.
\newblock Superconductivity of metals, taking into account the overlapping of
  energy bands.
\newblock {\em Phys. Met. Metallogr.}, 8(25), 1959.

\bibitem{Shanenko2011}
A.~A. Shanenko, M.~V. Milo\ifmmode \check{s}\else
  \v{s}\fi{}evi\ifmmode~\acute{c}\else \'{c}\fi{}, F.~M. Peeters, and A.~V.
  Vagov.
\newblock Extended {G}inzburg-{L}andau formalism for two-band superconductors.
\newblock {\em Phys. Rev. Lett.}, 106:047005, Jan 2011.

\bibitem{Vagov2012EGL}
A.~V. Vagov, A.~A. Shanenko, M.~V. Milo\ifmmode \check{s}\else
  \v{s}\fi{}evi\ifmmode~\acute{c}\else \'{c}\fi{}, V.~M. Axt, and F.~M.
  Peeters.
\newblock Extended {G}inzburg-{L}andau formalism: Systematic expansion in small
  deviation from the critical temperature.
\newblock {\em Phys. Rev. B}, 85:014502, Jan 2012.

\bibitem{Larkin}
A.~Larkin and A.~Varlamov.
\newblock {\em Theory of Fluctuations in Superconductors}.
\newblock Oxford University Press, 2005.

\bibitem{Gorkov1958}
L.~P. Gor'kov.
\newblock On the energy spectrum of superconductors.
\newblock {\em J. Exptl. Theoret. Phys.}, 34(7), 1958.

\bibitem{AGD1965}
A.~A. Abrikosov, I.~E. Dzyaloshinski, and L.~P. Gor'kov.
\newblock {\em Quantum Field Theoretical Methods in Statistical Physics}.
\newblock Pergamon, 1965.

\end{thebibliography}

\end{document}